\documentstyle{article}		

\newcommand\calv{{\cal V}}

\title{Radiative neutrino decay in hot media}
\author{Jos\'e F. Nieves\\
Laboratory of Theoretical Phsyics\\
Department of Physics, University of Puerto Rico\\
P.O. Box 23343, 
R\'{\i}o Piedras, Puerto Rico 00931-3343\\
\and
Palash B. Pal\\
Indian Institute of Astrophysics, Bangalore 560034, India}

\date{hep-ph/9702283}

\begin{document}

\maketitle 

\begin{abstract}     

We calculate the rate for the radiative neutrino decay in a thermal
background of electrons and photons, taking into account the effect of
the stimulated emission of photons in the thermal bath.  We show that
the rate is enhanced by a large factor relative to the rate for the
corresponding process in the vacuum.

\end{abstract}

The notion that the presence of an ambient medium modifies the
properties of elementary particles is now well known.  Sometimes the
effects are dramatic, as is the case of the MSW mechanism which has
been invoked to explain the solar neutrino puzzle \cite{msw}.  There, under
favorable conditions, the conversion of one neutrino flavor into
another is greatly enhanced by the presence of the background.  
Some time ago it was observed \cite{DNP89}
that the electromagnetic properties of neutrinos
are also drastically modified within a medium
when compared to the properties in vacuum.
This fact can lead to such interesting phenomena
as the Cherenkov radiation by massless neutrinos \cite{DNP96}
and the possible explanation of pulsar velocities by
neutrino oscillations biased by a magnetic field \cite{ks}.

It has also been shown previously \cite{DNP90,GKL91} 
that the rate for the radiative decay of
a massive neutrino that propagates through a
background of particles, 
	\begin{eqnarray} 
\nu (k) \to \nu'(k') + \gamma (q) \,, 
	\end{eqnarray}
can be much larger than the rate for the corresponding process in
the vacuum.  In the calculation of Ref.\ \cite{DNP90} the background
was taken to be a thermal bath of electrons, nucleons and photons, as
is the case of a plasma made of ordinary matter, as well as possibly
their antiparticles. Because the medium contains electrons but not
muons or tauons, the GIM mechanism that inhibits the radiative
neutrino decay in a vacuum is not operative anymore, and as a result
the rate can increase by a considerable amount.  It was also noted
there that the rate increases even further due the stimulated emission
of photons in the thermal background.  However, in those preliminary
calculations the effect just mentioned was neglected.

We have calculated the radiative decay rate including the effect of
the stimulated emission of photons.  In this note we present the
results of that calculation, and we show that, under the appropriate
conditions, the inclusion of this effect enhances the rate further by
many orders of magnitude.  

We follow the notations and conventions of Ref.~\cite{DNP90}. The
initial neutrino was assumed to have a mass $m$ and to be travelling
with a speed $\cal V$ in the medium, while the final neutrino $\nu'$
was assumed to be massless.  For all the three particles involved, we
disregarded the modification of the dispersion relations due to the
presence of the matter background.  As a consequence of this, only the
transverse modes of the photon contribute.  Then, retaining only the
background-induced contribution to the decay process, the square of
the transition matrix element was found to be
	\begin{eqnarray}
\left| {\cal M} \right|^2 = m^2 \left| U_{e\nu}^* U_{e\nu'} 
{\cal T}_T \right|^2 \left[ {(k+k') \cdot v \over \Omega} - 
{m^2 \over 2\Omega^2} \right] \,,
	\end{eqnarray}
where $v^\mu$ is the velocity four-vector of the background medium,
$\Omega=q\cdot v$, the elements of the matrix $U$ denote the mixing of
the neutrinos with the electron, and
	\begin{eqnarray}
{\cal T}_T = - 2\surd2 e G_F \int {d^3p \over (2\pi)^3 E} 
\left\{ f_-(p) + f_+(p) \right\} \,.
	\end{eqnarray}
Here $f_\pm$ stand for the positron and electron
distribution functions which, in terms of
the temperature $T=1/\beta$ and chemical
potential $\mu$, are given by 
	\begin{eqnarray}
f_\pm (p) = {1 \over \exp [\beta (p\cdot v \mp \mu)] + 1} \,,
	\end{eqnarray}
with
	\begin{eqnarray}
p^\mu \equiv (E, \vec p) \,, \quad E= \sqrt{\vec p \, ^2 + m_e^2}\,. 
	\end{eqnarray}

The differential decay rate can be written as
	\begin{eqnarray}\label{dGam}
d\Gamma' = {1\over 2k_0} (2\pi)^4 \delta^4 (k-k'-q) | {\cal M} |^2 
\times {d^3k' \over (2\pi)^3 2k'_0} {d^3q \over (2\pi)^3 2q_0} 
\left[ 1 + f(\Omega) \right] \,,
	\end{eqnarray}
where 
	\begin{eqnarray}
f(\Omega) = {1 \over e^{\beta\Omega} -1} \
	\end{eqnarray}
is the Planck distribution.  As already mentioned in
Ref.~\cite{DNP90}, the effect of the stimulated emission is taken into
account precisely by the factor $1+f(\Omega)$.  In the earlier
calculations it was set equal to unity, thus disregarding the effect
of stimulated emission. In that case, integrating Eq.\ (\ref{dGam}) in
the rest frame of the medium, where $v^\mu=(1, \vec 0)$ and
$\Omega=q_0$, we obtained
	\begin{eqnarray}\label{Gammanotstim}
\Gamma' = {m \over 16\pi} \left| U_{e\nu}^* U_{e\nu'} 
{\cal T}_T \right|^2 F(\calv) \,,
\label{Gam}
	\end{eqnarray}
where
\begin{equation}
\label{F}
F(\calv) = \sqrt{1- \calv^2} \left[ {2 \over \calv} \ln \left( {1+
\calv \over 1- \calv} \right) - 3 \right] \,.
\end{equation}
The result of including the effect of stimulated emission 
can be represented by 
replacing $F(\calv)$ in Eq.\ (\ref{Gam}) by the new function
	\begin{eqnarray}
F_T(\calv,\beta m) \equiv F(\calv) + F_s (\calv,\beta m) \,,
	\end{eqnarray}
where $F_s$ can be expressed as
\begin{eqnarray}
\label{Fs}
F_s(\calv,\beta m) & = & \left(\frac{1 - \calv^2}{2\calv}\right)
\int_{x_1}^{x_2}dx\; f(\Omega)
\left\{\frac{4}{x\sqrt{1 - \calv^2}} - \frac{2}{x^2} - 1\right\}
\end{eqnarray}
The variable $x$ is related to the photon energy by
\begin{equation}
x\equiv \frac{2\Omega}{m} \,,
\end{equation}
and the limits of integration are
\begin{equation}
x_1 = \frac{1}{x_2} = \sqrt{\frac{1 - \calv}{1 + \calv}}\,.
\end{equation}

The function $F(\calv)$ is given by a formula similar to Eq.\
(\ref{Fs}), but with the factor $f(\Omega)$ replaced by 1, which
yields the result quoted in Eq.\ (\ref{F}).  For arbitrary values of
the temperature and incident neutrino energy, $F_s$ can be evaluated
only numerically.  Analytical expressions, which show the general
features of the numerical results, can be obtained in some limiting
situations as we now discuss.

For $\calv = 0$ the photons emitted in the decay have an energy
$\Omega = m/2$, which is independent of the 
direction in which the photon is emitted.
The stimulated emission factor then becomes
\begin{equation}
f(\Omega) = \frac{1}{e^{\beta m/2} -1}\,,
\end{equation}
and therefore
it is simply a constant factor in the integration in Eq.\ (\ref{Fs}).
Thus, for slowly moving incident neutrinos, but any arbitrary
value of the temperature, 
\begin{eqnarray}
F_s(\calv,\beta m) \approx F_s(0,\beta m) & = & \frac{1}{e^{\beta m/2} -1}\\
 \label{FszeroVhightT}
& \approx &\frac{2}{\beta m}\,, \mbox{ (for $\beta m \ll 1$)}
\end{eqnarray}
where we have given, in Eq.\ (\ref{FszeroVhightT}), the limiting value
for high temperaures, as indicated.

However, the effect of stimulated emission is important for
temperatures that are much larger than the typical energy of the
photons emitted in the decay process, which is of the order of the
incident neutrino energy
\begin{equation}
k\cdot v = \frac{m}{\sqrt{1 - \calv^2}}\,.
\end{equation}
In this regime ($\beta k\cdot v \ll 1$), we can use the
approximation 
	\begin{eqnarray}
f(\Omega) \approx \frac{1}{\beta\Omega} \,,
	\end{eqnarray}
in which case the expression for $F_s$ can be evaluated analytically
for arbitrary values of the incident neutrino energy.
Introducing the function $f_s(\calv,\beta m)$ by writing
\begin{equation}\label{fs}
F_s (\calv,\beta m) \equiv \frac{2}{\beta m} f_s(\calv,\beta m)\,,
\end{equation}
the result of the integration for small $\beta m$ is
\begin{equation}\label{fszerobeta}
f_s(\calv,0) = 2 - \frac{1-\calv^2}{2\calv} \ln
\left( {1+ \calv \over 1-\calv} \right) \,.
\end{equation}
Notice, in particular, 
that Eq.\ (\ref{fszerobeta}) reproduces the result 
of Eq.\ (\ref{FszeroVhightT}),
$F_s(0,\beta m) = 2/\beta m$, as it should.
\begin{figure}
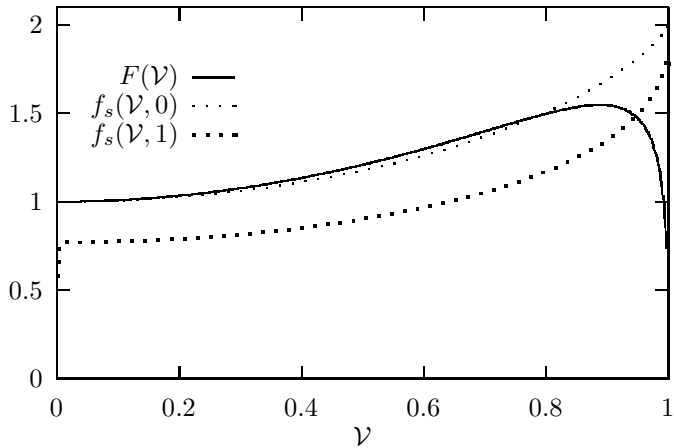

\begin{center}
\setlength{\unitlength}{0.240900pt}
\ifx\plotpoint\undefined\newsavebox{\plotpoint}\fi
\sbox{\plotpoint}{\rule[-0.200pt]{0.400pt}{0.400pt}}%

\end{center}
\caption[]{\label{fig:Fandfs}
Velocity dependence of the function $f_s(\calv,\beta m)$,
 defined in Eq.\ (\ref{fs}) of the text, for $\beta m=0$ and 
$\beta m =1$. Also shown is the velocity dependence of $F(\calv)$
defined in  Eq.\ (\ref{F}).}
\end{figure}
In Fig. \ref{fig:Fandfs} we have plotted $f_s(\calv,0)$, from which we
see that it is a very slowly increasing function that varies between 1
at $\calv=0$ to 2 at $\calv=1$. While $F(\calv)$ also is a slowly
varying function for most values of $\calv$, it drops sharply
\cite{GKL91} near $\calv=1$. However, $f_s(\calv,\beta m)$ does not
drop below the value it has at $\calv = 0$, for any value of the
temperature within the entire range of allowed values for
$\calv$. Moreover, it will have to remembered that, in the decay rate,
$f_s$ is multiplied by $2/\beta m$, which is a large number for
temperatures much higher than the mass of the decaying neutrino.  It
then follows that, for high temperatures, the factor $F + F_s$ is
dominated by the leading term of $F_s$ given by Eqs.\ (\ref{fs}) and
(\ref{fszerobeta}), and the rate is given approximately by
\begin{equation}
\label{Gammastim}
\Gamma^\prime\approx\Gamma_s\equiv
\frac{1}{8\pi\beta} \left| U_{e\nu}^* U_{e\nu'} 
{\cal T}_T \right|^2 f_s(\calv,0) \,, \mbox{ for $\beta m \ll 1$}\,.
\end{equation}
Comparing this with Eq.\ (\ref{Gammanotstim}), it follows that
there can be a considerable enhancement of the rate when the
temperature of the ambient photon gas is large compared to 
the mass of the decaying neutrino.

The formula in Eq.\ (\ref{Gammastim}) has another peculiar
character. Notice that, although it has a dependence on the mass
matrix of the neutrinos through the elements of the mixing matrix, the
decay rate does not directly depend on the mass of the decaying
neutrino. Thus, irrespective of the mass, any neutrino will have the
same decay rate apart from the difference coming from the mixing
angles. To estimate this rate, we recall \cite{DNP90} that if the
background medium consists of non-relativistic electrons, 
	\begin{eqnarray}
{\cal T}_T^{\rm (NR)} = - \surd 2 e G_F n_e/m_e \,,
	\end{eqnarray}
where $n_e$ is the number density of the electrons. On the other hand,
for a background of extremely relativistic electrons
and positrons,
assuming that the electron and photon temperatures are equal
and the chemical potential is zero, one obtains
	\begin{eqnarray}
{\cal T}_T^{\rm (ER)} = -  {e G_F \over 3 \surd 2 \beta^2} \,.
	\end{eqnarray}
Inserting these expressions into Eq.\ (\ref{Gammastim}),
we obtain the decay rate
\begin{eqnarray}
\Gamma^{\prime(\rm NR)} &\approx& \alpha G_F^2 
\left| U_{e\nu}^* U_{e\nu'}\right|^2 f_s(\calv,0)
\frac{n_e^2}{m_e^2\beta} \nonumber\\*
&=& (4\times 10^{18} {\rm s})^{-1} 
\left| U_{e\nu}^* U_{e\nu'}\right|^2 f_s(\calv,0)
\left( {T \over 1 \; {\rm MeV}} \right) 
\left( {n_e \over 10^{24} \; {\rm cm}^{-3}} \right)^2 
\label{GammaNR}
\end{eqnarray}
for a background of non-relativistic electrons, and
\begin{eqnarray}
\Gamma^{\prime(\rm ER)} &\approx& \alpha G_F^2 
\left| U_{e\nu}^* U_{e\nu'}\right|^2 f_s(\calv,0)
\frac{1}{36\beta^5} \nonumber\\* 
&=& (2.5\times 10^4 {\rm s})^{-1} 
\left| U_{e\nu}^* U_{e\nu'}\right|^2 f_s(\calv,0)
\left( {T \over 1 \; {\rm MeV}} \right)^5 
\label{GammaER}
\end{eqnarray}
for an extremely relativistic electron gas at zero chemical 
potential.

The quantitative study of the electromagnetic
properties of neutrinos in a background of particles
can lead to interesting physical phenomena
and important applications.  As already mentioned,
the phenomenon of neutrino conversion in magnetized media
has opened new possibilities for explaining
the peculiar velocity of pulsars \cite{ks}.
The results that have been presented
in the present work can also have relevance
in cosmological and astrophysical contexts such
as the early universe and the supernova, 
as well as in laboratory experiments related to
rare processes and the coherent pehenomena
in crystals, where possible applications
to the observation of induced radiative neutrino
transitions have been mentioned \cite{zoller}.

The research of JFN was 
supported in part by
the US National Science Foundation Grant PHY-9600924.

\end{document}